\documentclass[twocolumn,showpacs,preprintnumbers]{revtex4}
\usepackage{amsfonts}
\usepackage{amsmath}

\normalbaselines
\input{tcilatex}

\begin{document}

\author{F.J. L\'{o}pez-Rodr\'{\i}guez$^{1}$, G.G. Naumis$^{1,2,3}$}
\affiliation{$^{1}$ Departamento de F\'{\i}sica-Qu\'{\i}mica, Instituto de F\'{\i}sica,
Universidad Nacional Aut\'{o}noma de M\'{e}xico (UNAM) Apartado Postal
20-364, 01000, M\'{e}xico D.F., MEXICO. }
\affiliation{$^{2}$Facultad de Ciencias, Universidad Aut\'{o}noma del Estado de Morelos,
Av. Universidad 10001, 62210, Cuernavaca, Morelos, Mexico }
\affiliation{$^{3}$Deptartamento de F\'{\i}sica-Matem\'{a}tica, Universidad
Iberoamericana, Prolongaci\'{o}n Paseo de la Reforma 880, Col. Lomas de
Santa Fe, 01210, M\'{e}xico DF, Mexico.}

\begin{abstract}
We find the exact solution of graphene's carriers under electromagnetic
radiation. To obtain such solution, we combine Floquet theory with a trial
solution. Then the energy spectrum is obtained without using any
approximation. Using such results, we prove that the energy spectrum
presents a gap opening which depends on the radiation frequency and electric
wave intensity, whereas the current shows a strongly non-linear behaviour.
\end{abstract}

\date{\today }
\title{Electrons and holes in graphene under electromagnetic waves: gap
appearance and non-linear effects}
\maketitle

\bigskip

Graphene, a two dimensional allotrope of carbon, has been the centre of a
lot of research since its experimental discovery four years ago \cite%
{Novoselov}. It has amazing properties \cite{Katsnelson}\cite{Kim}\cite%
{NovoGeim}. For instance, electrons in graphene behave as massless
relativistic fermions \cite{Semenoff},\cite{NovoselovGeim} . Such property
is a consequence of its bipartite crystal structure \cite{Weiss}, in which a
conical dispersion relation appears near the K, K' points of the first
Brioullin zone\cite{Wallace}. Among other properties one can cite the high
mobility that remains higher even at high electric-field induced
concentration and translates into ballistic transport on a submicron scale 
\cite{Geim} at $300%
{{}^\circ}%
K$. Graphene is therefore a promising material for building electronic
devices, but there are some obstacles to overcome. One is the transmission
probability of electrons in graphene, which can be unity irrespective of the
height and width of a given potential barrier. As a result, conductivity
cannot be changed by an external gate voltage, a feature required to build a
field-effect transistor (FET), although a quantum dot can be used \cite%
{Novoselov2}. In a previous paper, we have shown that a possible way to
induce a pseudogap around the Fermi energy consists in doping graphene\cite%
{naumis} . On the other hand, much efforts have been devoted to understand
the electrodynamic properties of graphene as well as its frequency dependant
conductivity \cite{peres}, \cite{Sharapov1},\cite{Sharapov2},\cite{Mikhailov}%
. In this Rapid Communication we solve, the problem of graphene's electron
behavior in the presence of an electromagnetic plane wave. As a result, in
the limit of long wave length of the electromagnetic field, we are able to
find a gap opening. This\ is like if electrons in graphene acquire an
effective mass under electromagnetic radiation. We also calculate the
current and show that there is in fact a strongly non-linear electromagnetic
response, as was claimed before by S. A. Mikhailov\cite{Mikailov1} using a
semi-classical approximation.\ 

\FRAME{ftbpFU}{3.0378in}{2.4815in}{0pt}{\Qcb{(Color online). Graphene
lattice. Atoms in the A sublattice are shown with different color than those
in the B sublattice. The vector $\protect\overrightarrow{k}$ in which the
electromagnetic field propagates and the electric field vector $\protect%
\overrightarrow{E\text{ }}$ are shown.}}{\Qlb{figura1}}{mejor dibujo_7.eps}{%
\special{language "Scientific Word";type "GRAPHIC";display
"USEDEF";valid_file "F";width 3.0378in;height 2.4815in;depth
0pt;original-width 10.0024in;original-height 7.4936in;cropleft "0";croptop
"1";cropright "1";cropbottom "0";filename 'Mejor
dibujo_7.eps';file-properties "XNPEU";}}

Considering an electron in a graphene lattice subject to an electromagnetic
plane wave as shown in Fig. 1. The plane wave propagates along the
two-dimensional space where electrons move. In this Rapid Communication we
take $\mathbf{k}=(0,k)$. The generalization to any direction of $\mathbf{k}$
is straightforward. It has been proved that a single-particle Dirac
Hamiltonian can be used as very good approximation to describe charge
carriers dynamics in graphene\cite{Sarma}. For wave vectors close to the K
point of the first Brioullin zone, the Hamiltonian is \cite{Katsnelson2},$\
\ \ \ \ \ \ \ $%
\begin{equation}
H(x,y,t)=v_{F}\left( 
\begin{array}{cc}
0 & \hat{\pi}_{x}-i\hat{\pi}_{y} \\ 
\hat{\pi}_{x}+i\hat{\pi}_{y} & 0%
\end{array}%
\right) ,  \label{hamiltonian}
\end{equation}%
where $v_{F}$ is the Fermi velocity $v_{F}\approx c/300$, \ $\hat{\pi}=%
\mathbf{\hat{p}}-e\mathbf{A}/c$ with\textbf{\ }$\mathbf{\hat{p}}$\textbf{\ }%
being the electron momentum operator and $\mathbf{A}$\textbf{\ \ }is the
vector potential of the applied electromagnetic field, given by $\mathbf{A=(}%
\frac{E_{0}}{\omega }\cos (ky-\omega t),0)$ , where $E_{0}$ is the amplitude
of the electric field and $\omega $ is the frequency of the wave. $E_{0\text{
}}$is taken as a constant since screening effects are weak in graphene \cite%
{DiVicenzo}. For the valley K' the signs before $\pi _{y}$ are the opposite 
\cite{Katsnelson2}. The dynamics is governed by

\begin{equation}
H(x,y,t)\mathbf{\Psi }(x,y,t)=i\hbar \frac{\partial \mathbf{\Psi }(x,y,t)}{%
\partial t},  \label{Dirac equation}
\end{equation}%
where 
\begin{equation*}
\mathbf{\Psi }(x,y,t)=\left( 
\begin{array}{c}
\Psi _{A}(x,y,t) \\ 
\Psi _{B}(x,y,t)%
\end{array}%
\right)
\end{equation*}%
is a two component spinor. $\ $Here $A$ and $B$ stand for each sublattice
index of the bipartite graphene lattice \cite{Weiss}.To find the eigenstates
and eigenenergies we adapt a method developed by D. M. Volkov in 1935 \cite%
{Berestetskii} to study the movement of relativistic particles under an
electromagnetic field. Volkov considered the complete four component
bispinor and Dirac matrices. We use instead only Pauli matrices, and two
component spinors. First we write the equations of motion for each component
of the spinor

$\ \ \ \ \ \ \ \ \ \ \ \ \ \ \ \ \ \ \ \ \ $%
\begin{equation}
\ v_{F}\ (\hat{\pi}_{x}-i\hat{\pi}_{y})\Psi _{B}(x,y,t)=i\hbar \frac{%
\partial \Psi _{A}(x,y,t)}{\partial t},  \label{ecuacion1}
\end{equation}

\begin{equation}
\ v_{F}(\hat{\pi}_{x}+i\hat{\pi}_{y})\Psi _{A}(x,y,t)=i\hbar \frac{\partial
\Psi _{B}(x,y,t)}{\partial t},  \label{ecuacion2}
\end{equation}%
Considering the magnetic field as $\mathbb{B}$, the commutation rules for $%
\hat{\pi}_{x}$ and $\hat{\pi}_{y}$ are in this case,

\begin{equation}
\left[ \hat{\pi}_{i},\hat{\pi}_{j}\right] =\frac{i\hbar e}{c}\varepsilon
_{ijk}\mathbb{B}_{k},\ \ \ \ i,j=x,y  \label{conmutacion1}
\end{equation}%
\ \ $\ \ \ \ \ \ \ \ \ $%
\begin{equation}
\left[ \frac{\partial }{\partial t},\hat{\pi}_{x}\pm i\hat{\pi}_{y}\right] =-%
\frac{eE_{o}}{c}\sin (ky-\omega t),  \label{conmutacion2}
\end{equation}%
and using that $k_{\mu }A^{\mu }=0$, we find the following equation of
motion for the spinor:

\begin{multline}
-\hbar ^{2}\left( v_{F}^{2}\left( \frac{\partial ^{2}\mathbf{\Psi }}{%
\partial x^{2}}+\frac{\partial ^{2}\mathbf{\Psi }}{\partial y^{2}}\right) -%
\frac{\partial ^{2}\mathbf{\Psi }}{\partial t^{2}}\right) \\
+2i\hbar \xi v_{F}\cos \phi \frac{\partial \mathbf{\Psi }}{\partial x}
\label{Ecuacion completa1} \\
+\left[ \xi ^{2}\cos ^{2}\phi \mathbf{-}\xi v_{F}\hbar \sigma _{z}k\sin \phi 
\mathbf{-}i\hbar \omega \xi \sigma _{x}\sin \phi \right] \mathbf{\Psi =0,}
\end{multline}%
where we have defined the phase $\phi $ of the electromagnetic wave as $\phi
=ky-\omega t$. The parameter $\xi $ is defined $\xi =\frac{eE_{0}v_{F}}{%
c\omega }$, and $\sigma _{\mu }$ is the set of Pauli matrices. To solve this
equation, we follow Volkov%
\'{}%
s suggestion to use a trial function that depends upon the phase $\phi $ of
the wave with the following form,\ \ \ \ \ \ \ \ \ \ \ \ \ \ \ \ \ \ \ \ \ \
\ \ \ \ \ \ \ \ \ \ \ \ \ \ \ \ \ \ \ \ \ \ \ \ \ \ \ \ \ \ \ \ \ \ \ \ \ \
\ \ \ \ \ \ \ \ \ \ \ \ \ \ \ \ \ \ \ \ \ \ \ \ \ \ \ \ \ \ \ \ \ \ \ \ 

\begin{equation}
\mathbf{\Psi }(x,y,t)=e^{ip_{x}x/\hbar +ip_{y}y/\hbar -i\epsilon t/\hbar }%
\mathbf{F}(\phi ),  \label{Solucion1}
\end{equation}%
where $\epsilon =v_{F}\sqrt{p_{x}^{2}+p_{y}^{2}}$ and $\mathbf{F}(\phi )$ is
a spinor. Inserting Eq. (\ref{Solucion1}) into Eq.( \ref{Ecuacion completa1}%
), it yields a equation for $\mathbf{F}(\phi )$. The resulting differential
equation is,

\begin{multline}
-\hbar ^{2}(v_{F}^{2}k^{2}-\omega ^{2})\frac{d^{2}\mathbf{F}(\phi )}{d\phi
^{2}}+2i\hbar \epsilon \eta \frac{d\mathbf{F}(\phi )}{d\phi }\mathbf{+}%
[-2\xi v_{F}p_{x}\cos \phi -\xi v_{F}\hbar \sigma _{z}k\sin \phi 
\label{ecuacion F} \\
+\xi ^{2}\cos ^{2}\phi -i\hbar \omega \xi \sigma _{x}\sin \phi ]\mathbf{F}%
(\phi )=0,
\end{multline}%
and $\eta =$ $\epsilon \omega -v_{F}^{2}kp_{y}$. The solution to previous
equation has the form $\mathbf{F}(\phi )=\exp (i\phi B/2A)z(\phi )$, where $%
z(\phi )$ is the solution of

\begin{equation*}
z^{\prime \prime }(\phi )+[(C(\phi )/A)-B^{2}/(2A^{2})]z(\phi )=0,
\end{equation*}%
$A=-\hbar ^{2}(v_{F}^{2}k^{2}-\omega ^{2})$, $B=2\hbar \eta $ and $C(\phi
)=-2\xi v_{F}p_{x}\cos \phi -\xi v_{F}\hbar \sigma _{z}k\sin \phi +\xi
^{2}\cos ^{2}\phi -i\hbar \omega \xi \sigma _{x}\sin \phi .$ On the other
hand this equation can also be solved if we consider the long wave limit ($%
\lambda \rightarrow \infty $) and if it is neglected the term with second
derivative upon $\phi $ to give,

\begin{equation}
\mathbf{F}(t)=\exp \left[ G(t)\right] \mathbf{u},  \label{solucion3}
\end{equation}%
where $\mathbf{u}$ is a two component spinor and%
\begin{multline}
G(t)=-\frac{i\xi ^{2}}{4\hbar \epsilon }t-\frac{i\xi v_{F}}{\hbar \epsilon
\omega }\sin \omega tp_{x}  \label{ecuacion12} \\
+\frac{i\xi ^{2}}{8\hbar \epsilon \omega }\sin 2\omega t-\frac{\xi }{%
2\epsilon }\sigma _{x}\cos \omega t\text{,}
\end{multline}

The complete solution to our problem is:

\begin{equation}
\mathbf{\Psi (x,y,t)=}\exp \left[ ip_{x}x/\hbar +ip_{y}y/\hbar -i\epsilon
t/\hbar \right] \exp \left[ G(t)\right] \mathbf{u,}  \label{solucioncompleta}
\end{equation}%
where\textbf{\ }$\mathbf{u}$ must be taken as a two component spinor which
satisfies the requirement that when $\mathbf{A}\rightarrow 0$, $\mathbf{\Psi 
}(x,y,t)$\textbf{\ }should be the solution of \ the free Dirac equation to
avoid strange solutions. Thus $\mathbf{u}$ is given by:

\begin{equation}
\mathbf{u}=\frac{1}{\sqrt{2}}\left( 
\begin{array}{c}
\pm e^{-i\varphi /2} \\ 
e^{i\varphi /2}%
\end{array}%
\right) \text{, \ \ \ \ \ \ \ \ \ \ \ }\varphi =\tan (\frac{p_{y}}{p_{x}})%
\text{ }  \label{espinor1}
\end{equation}%
where the sign plus stands for electrons and minus sign stands for holes.The
eigenvalues are $\epsilon +\xi ^{2}/4\epsilon $. But now we need to take
into account the time periodicity of the electromagnetic field, which means
that the solution must be written in the following way \cite{floquet},

\begin{equation}
\mathbf{\Psi }=\exp (-i\chi t/\hbar )\mathbf{\Phi }(x,y,t),
\label{floquet forma}
\end{equation}%
where $\mathbf{\Phi }(x,y,t)$ is periodic in time, \textit{i .e. }$\mathbf{%
\Phi }(x,y,t)=\mathbf{\Phi }(x,y,t+T)$, and $\chi $ is a real parameter,
being unique up to multiples of $\hbar \omega $, $\omega =2\pi /\tau $. In
fact, such property is very well known within the so-called Floquet theory 
\cite{floquet}, developed for time dependent fields. This time periodicity
of the field leads to the formation of bands. Our solution Eq.(\ref%
{solucioncompleta}) satisfies the requisite form Eq. (\ref{floquet forma}).
However, to obtain the Floquet states \cite{floquet} for this problem, we
need to consider that $\epsilon +\xi ^{2}/4\epsilon $ is unique up to
multiples of $\hbar \omega .$ Thus $\ E=\epsilon +\xi ^{2}/4\epsilon +$ $%
n\hbar \omega $, where $n$ is an integer such that $n=0,\pm 1,\pm 2,...$,
while $\mathbf{\Phi (}x,y,t)$ is \ the Floquet mode given by, 
\begin{equation}
\mathbf{\Phi (}x,y,t)=\exp \left[ in\omega t+ip_{x}x/\hbar +ip_{y}y/\hbar %
\right] \exp \left[ F(t)\right] \mathbf{u.}  \label{floquet mode}
\end{equation}%
where $F(t)=G(t)+i\xi ^{2}t/4\hbar \epsilon .$The Floquet \ modes of the
problem must satisfy the equation \cite{floquet},%
\begin{equation}
\left[ H(x,y,t)-i\hbar \frac{\partial }{\partial t}\right] \mathbf{\Phi (}%
x,y,t)=E\mathbf{\Phi (}x,y,t)
\end{equation}

\bigskip The final eigenenergies for electron and holes are the following,

\begin{equation}
E_{n}(p)=n\hbar \omega \pm v_{F}\left\Vert p\right\Vert \pm \frac{%
e^{2}E_{0}^{2}v_{F}}{4c^{2}\omega ^{2}\left\Vert p\right\Vert },\text{ \ \ }
\label{energias}
\end{equation}%
where $n=0,\pm 1,...$, and the wave-function is,

\begin{equation}
\mathbf{\Psi }_{n,p}(x,y,t)=\exp (-iE_{n}(p)t/\hbar )\mathbf{\Phi (}x,y,t),
\label{solucionperiodica1}
\end{equation}%
\bigskip and $\mathbf{\Phi (}x,y,t)$ is the Floquet mode Eq.( \ref{floquet
mode}).I f the two smallest terms are neglected in Eq.( \ref{ecuacion12}),
our solution can be reduced to one found in Ref. \cite{Trauzettel} . The
spectrum given by Eq. (\ref{energias}) is made of bands, where the
electromagnetic field bends the linear dispersion relationship due to the
last term.

\begin{equation}
\Delta (p)=\frac{e^{2}E_{0}^{2}v_{F}}{4c^{2}\omega ^{2}\left\Vert
p\right\Vert }=\frac{\xi ^{2}}{4\epsilon }  \label{gap}
\end{equation}%
Thus, around the Fermi energy, holes and electron bands are separated by a
gap of size $\Delta (p)=2\Delta (p_{F})$. To estimate the magnitude of the
gap, we consider electrons near the Fermi energy, thus $\epsilon
=v_{F}\left\Vert p\right\Vert \approx \epsilon _{F}$, where $\epsilon _{F}$
is approximately \cite{Sharapov1} $\epsilon _{F}=86meV$. \ For a typical
microwave frequency $\omega =50GHz$ with an intensity $E_{0}=3V/cm$ of the
electric field, the gap size is around $\Delta \approx 0.2meV$ .\ Due to the
gap opening, the particles are not longer massless. The mass acquired by the
carriers due to the electromagnetic field is therefore around $10^{-4}m_{e}$%
. Since this is a time-dependant problem, one cannot measure the gap
directly from the density of states. Instead, one can look for jumps in the
conductance, using, for example, the device proposed in Ref .\cite%
{Trauzettel}. Finally, the solution can be used to evaluate the current
using the collisioness Boltzmann equation\cite{Mikailov1}, neglecting
interband transitions.However, this approach \ turns out to give the same
results as considering the velocity of the particles as time dependant while
the distribution function [$f_{p}/t)$] remains static\cite{Torres}. Such
equivalence is the result of the trivial statement that in the Boltzmann
approach, the electric field induces a displacement fo the Fermi surface\cite%
{Torres}.In fact, the current obtained below is equal to the one obtained
using a Boltzmann equation semiclassical approach for graphene under an
electric adiabatic field \cite{Mikailov1}, which also reproduces the
intraband Drude conductivity\cite{Mikailov1}. The electric current is given
by $\mathbf{j}(t)=4S^{-1}\sum_{\mathbf{p}}\mathbf{j}_{\mathbf{p}}\mathbf{(}%
t)f_{\mathbf{p}}(t)$, where $S^{-1}$ is the sample area and the factor $4$
comes for the spin and valley degeneracy. $\mathbf{j}_{\mathbf{p}}\mathbf{(}%
t)$ denotes the contribution to the current of particles with momentum $%
\mathbf{p}$ at time $t$, while $f_{\mathbf{p}}(t)$ is the distribution
function in phase space at temperature $T$. Let us first calculate the $\mu $
component of the current vector, given by $j_{\mu ,\mathbf{p}}\mathbf{(}%
t)=ev_{F}\mathbf{\Psi }_{n,p}^{\ast }\sigma _{\mu }\mathbf{\Psi }_{n,p}$. We
use Eq. (\ref{solucionperiodica1}) to obtain the components of $j$ in the $x 
$ and $y$ directions,

\begin{equation}
j_{x,p}(t)=ev_{F}\sinh \left( \frac{\xi }{\epsilon }\cos \omega t\right)
+\cosh \left( \frac{\xi }{\epsilon }\cos \omega t\right) \cos \varphi ,
\end{equation}%
and $j_{y,p}=ev_{F}\sin \varphi $. In $j_{x}$ we already observe a very
important non-linear behavior. In fact, $j_{x,p}(t)$ can be written as a
combination of harmonics by using a Fourier series development of the
hyperbolic functions \cite{Ryzhik},%
\begin{equation*}
j_{x,p}(t)=ev_{F}\dsum\limits_{s=0}^{\infty }J_{2s+1}\left( \frac{\xi }{%
\epsilon }\right) \cos ((2s+1)\omega t)
\end{equation*}

\begin{equation}
+ev_{F}\left( J_{0}\left( \frac{\xi }{\epsilon }\right)
+2\dsum\limits_{s=1}^{\infty }J_{2s}\left( \frac{\xi }{\epsilon }\right)
\cos (2s\omega t)\right) \cos \varphi ,  \label{Bessel}
\end{equation}%
where $J_{s}(\xi /\epsilon )$ is a Bessel function. Our result is similar to
the first terms of the current obtained using a semi-classical approximation 
\cite{Mikailov1}, except for the last term which eventually cancels out in
the thermodynamical limit. Now we include such limit by using $f_{\mathbf{p}%
}(t)$. In this case, we are dealing with quasiparticles with an effective
dispersion relation. Using Eq. (\ref{Bessel}), and by summing over the phase
space we get,%
\begin{equation}
j_{x}(t)=\frac{4ev_{F}}{(2\pi \hbar )^{2}}\dsum\limits_{s=0}^{\infty
}A(s)\cos ((2s+1)\omega t),
\end{equation}%
\begin{equation*}
A(s)\equiv \dint\limits_{-\infty }^{\infty }\dint\limits_{-\infty }^{\infty
}J_{2s+1}\left( \frac{\xi }{v_{F}p}\right) n(p)dp_{x}dp_{y},
\end{equation*}%
where $n(p)=\left[ 1+\exp (E_{n}(p)-\epsilon _{F})/k_{B}T\right] ^{-1}$ is
the occupation factor. For $\epsilon _{F}\gg k_{B}T$, $n(p)$ can be replaced
by a step function. Using Eq. (\ref{energias}), $v_{F}p_{F}=\epsilon _{f}(1+%
\sqrt{1-(\xi /\epsilon _{f})^{2}})/2$ for $n=0$, $A(s)$ is,%
\begin{equation}
A(s)=2\pi p_{F}^{2}\alpha ^{2}\dint\limits_{0}^{1}J_{2s+1}\left( \frac{Q_{0}%
}{\alpha x}\right) xdx,
\end{equation}%
\begin{equation*}
\alpha \equiv \left( \frac{(1+\sqrt{1-Q_{0}^{2}})}{2}\right) \approx \left\{
1-\frac{1}{4}Q_{0}^{2}\right\} .
\end{equation*}%
where $Q_{0}=\xi /\epsilon _{F}=eE_{0}/(c\omega p_{F})$. For $s=0$ we obtain,

\begin{equation}
A(0)\approx\pi p_{F}^{2}\alpha Q_{0}\left( 1+\frac{1}{8}\left( \frac{Q_{0}}{%
\alpha}\right) ^{2}-\frac{1}{576}\left( \frac{Q_{0}}{\alpha}\right)
^{4}+...\right) ,
\end{equation}
and for $s>0$, the integral can be approximated as $A(s)\approx2\pi
p_{F}^{2}\left( Q_{0}/2\right) ^{3}\Gamma(s-\frac{1}{2})/\alpha\Gamma(s+%
\frac {5}{2})$. The first terms of the current are,

\begin{equation*}
j_{x}(t)\approx en_{e}v_{F}Q_{0}\left\{ 
\begin{array}{c}
\left( 1-\frac{1}{8}Q_{0}^{2}\right) \cos (\omega t) \\ 
+\frac{2}{15}Q_{0}^{2}\cos (3\omega t)+..%
\end{array}%
\right\} \text{,}
\end{equation*}%
where $n_{e}$ is the density of electrons $n_{e}=p_{F}^{2}/\pi \hbar ^{2}$.
\ In conclusion, we solved the Dirac equation for graphene charge carriers
under an electromagnetic field.

We acknowledge \ DGAPA-UNAM projects No. IN-117806, No. IN-111906 and
CONACyT Grants No. 48783-F and No. 50368.

\end{document}